# Quantum computers can search arbitrarily large databases by a single query

Lov K. Grover, 3C-404A Bell Labs, 600 Mountain Avenue, Murray Hill NJ 07974 *(lkgrover@bell-labs.com)*

**Summary**

This paper shows that a quantum mechanical algorithm that can query information relating to multiple items of the database, can search a database for a unique item satisfying a given condition, in a single query (a query is defined as any question to the database to which the database has to return a (YES/NO) answer). A classical algorithm will be limited to the information theoretic bound of at least $\log_2 N$ queries, which it would achieve by using a binary search.

**0. Background** Imagine the following situation: there are $N$ items in a database (say $A_1, A_2...A_N$). One of the items is marked. An oracle knows which item is marked; however, the oracle only gives one bit (YES/NO) answers to *any* questions that are posed to it. The challenge is to find out which item is marked with the minimum number of questions to the oracle. It is well known that the optimal way is to ask questions which eliminate half the items under consideration with each question - this process is known to computer scientists as a binary search and yields the answer after approximately $\log_2 N$ queries [Binary].

Quantum mechanical computers can be in a superposition of states and carry out multiple operations at the same time. An algorithm that uses this parallelism is [Search] which searches an $N$ item database for a single marked item in $O(\sqrt{N})$ quantum queries where each query pertains to only one of the $N$ items. This was in some ways a surprising result, in some ways not so surprising. To those familiar with classical entities, this was surprising since there are $N$ items to be searched, so how could the result be obtained in fewer than $N$ steps? However, from a quantum mechanical point of view all $N$ items are being simultaneously searched, so there is no obvious reason the results could not be obtained in a single query. By means of subtle reasoning about unitary transformations, [BBBV] & [BBHT] show that quantum mechanical algorithms cannot search faster than $\Omega(\sqrt{N})$ queries.

**1. This paper** This paper shows that in case it is possible to query the quantum computer about multiple items, then it is possible to search the entire database in a single query. In contrast, a classical computer will be limited to the information theoretic bound of $\log_2 N$ queries. However, the query is complicated and preparing the query and processing the results of the query take $\Omega(N \log N)$ steps.[1]

The algorithm works by considering a quantum system composed of multiple subsystems; each subsystem has an $N$ dimensional state space like the one used in the $O(\sqrt{N})$ quantum search algorithm [Search], i.e. each basis state of a subsystem corresponds to an item in the database. It is shown that with a *single* quantum query, pertaining to information regarding all $N$ items, the amplitude (and thus probability) in the state corresponding to the marked item(s) of *each* subsystem can be amplified by a small amount. By choosing the number of subsystems to be appropriately large, this small difference in probabilities can be estimated by making a measurement to determine which item of the database each subsystem corresponds to - the item pointed to by the most subsystems is the marked item.

A similar result has independently been obtained by Terhal & Smolin [Superfast] by a different approach.

**2. Inversion about average** Assume that there is a binary function $f(\bar{x})$ that is either $0$ or $1$. Given a superposition over states $\bar{x}$, it is possible to design a quantum circuit that will selectively invert the amplitudes in all states where $f(\bar{x}) = 1$. This is achieved by appending an ancilla bit, $b$ and considering the quantum circuit that transforms a state $|\bar{x}, b\rangle$ into $|\bar{x}, f(\bar{x}) XOR\, b\rangle$ (such a circuit exists since, as proved in [Reversible], it is possible to design a quantum mechanical circuit to evaluate any function $f(\bar{x})$ that can be evaluated classically). If the bit $b$ is initially placed in a superposition $\frac{1}{\sqrt{2}}(|0\rangle - |1\rangle)$, this circuit will invert the amplitudes precisely in the states for which $f(\bar{x}) = 1$, while leaving amplitudes in other states unchanged [BBHT].

By using such a selective inversion followed by an *inversion about average* operation, [Search] showed that the magnitude of the amplitude in marked state(s) can be increased by a certain amount. The *inversion about average* operation is defined by the following unitary operation $D$: $D_{ij} = \frac{2}{N}$ if $i \neq j$ & $D_{ii} = -1 + \frac{2}{N}$.

---

1. $O(f(x))$ means asymptotically *less* than a constant times $f(x)$,
   $\Omega(f(x))$ means asymptotically *greater* than a constant times $f(x)$.

Assume that $D$ is applied to a superposition with each component of the superposition, except one, having an amplitude equal to $\frac{1}{\sqrt{N}}$; the one component that is different has an amplitude of $-\frac{1}{\sqrt{N}}$. The one that was negative, now becomes positive and its magnitude increases to approximately $\frac{3}{\sqrt{N}}$, the rest stay virtually unchanged.

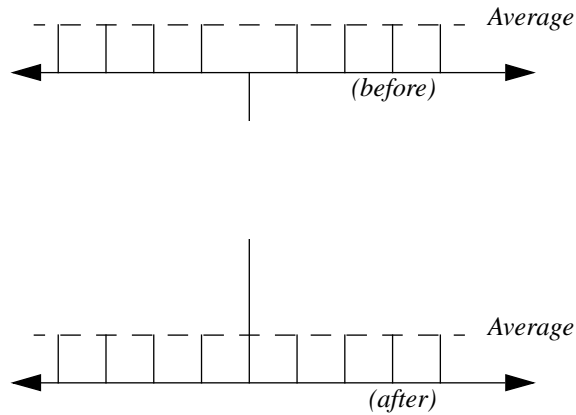

**The *inversion about average* operation is applied to a superposition in which all but one of the components is initially $\frac{1}{\sqrt{N}}$; one of the components is initially $-\frac{1}{\sqrt{N}}$.**

## 3. Algorithm with queries pertaining to multiple items

As mentioned in section 1, the algorithm assumes a large number of identical subsystems. Each subsystem has a basis state corresponding to an item of the database and it is placed in a superposition of these states. The aim is to boost the amplitude, and hence probability, of the basis state(s) corresponding to the marked item(s) in each subsystem by a small amount. If the number of subsystems is sufficiently large, then by carrying out an observation it is possible to infer, what basis state the probability is larger in and hence which basis state the amplitude has been boosted in and from this the marked item in the database. It is explained after step (iv) in this section, that the number of subsystems needs to be $\Omega(N\log N)$.

The algorithm is given below for a single marked item. A similar algorithm (and similar proof) work if multiple items are marked.

(i) Consider a tensor product of $\eta$ identical quantum mechanical subsystems - all subsystems have an $N$ dimensional state space. Each of the $N$ basis states corresponds to an item in the database. All $\eta$ subsystems are placed in a superposition with equal amplitude in all $N$ states.

*Assuming $N$ to be a power of 2, the state of each subsystem is initialized by taking a set of $\log_2 N$ qubits which gives $N$ states; the system consists of $\eta$ such subsystems. Each qubit is placed in the superposition $\frac{1}{\sqrt{2}}(|0\rangle + |1\rangle)$, thus obtaining equal amplitudes in all $N$ states. Denoting the $N$ states by $S_1, S_2....S_N$, the state vector is proportional to $(|S_1\rangle + |S_2\rangle + ... + |S_N\rangle)^\eta$ which may be written as $(|S_1 S_1 ... S_1\rangle + |S_1 S_1 ... S_2\rangle + ... + N^\eta$ such terms).*

(ii) Query the database as to whether the number of subsystems (out of the $\eta$ subsystems) in the state corresponding to the marked item, is odd or even. In case it is odd, invert the phase; if it is even, do nothing. This is achieved by using the technique described in section 2 with the function $f(\bar{x})$ equal to $1$ if the query indicates that the marked item's basis state was present an odd number of times, $0$ otherwise.

*Let $S_1$ be the state corresponding to the marked item. The state vector after this operation becomes: $(\pm |S_1 S_1 ... S_1\rangle \pm |S_1 S_1 ... S_2\rangle \pm ... N^\eta$ such terms), the sign of each term is determined by whether the state corresponding to the marked item ($S_1$) is present an odd or even number of times in the respective term. This state vector can be factored and written as: $(-|S_1\rangle + |S_2\rangle + ... + |S_N\rangle)^\eta$. The system is now in a tensor product of $\eta$ identical quantum mechanical subsystems, each of which has an $N$ dimensional state space; in each of the $\eta$ subsystems, the phase of the amplitude in the basis state corresponding to the marked item is inverted.*

*Note that by a single operation on the multisystem wavefunction, the wavefunction of each subsystem has been altered in a suitable way. Using a single query, the phase of the amplitude in the state corresponding to the marked item in each of these $\eta$ subsystems is inverted - the reason it needs only a single query is that the new phase can have only two possible values $(\pm 1)$, therefore the only statistic needed from the oracle is: "Is the number of subsystems in the state corresponding to the marked item is odd or even?"*

(iii)    Do a single inversion about average operation on each of the $\eta$ subsystems separately.

*Since the system is in a tensor product of $\eta$ identical quantum mechanical subsystem, each subsystem can be independently operated on. As mentioned at the end of section 2, if the magnitude of the amplitude in all states be equal, but the sign of the amplitude in one state be opposite, then the magnitude of the amplitude in the state with the negative amplitude can be increased by a factor of 3 by an inversion about average operation. The state vector after carrying out this operation becomes approximately: $(3|S_1\rangle + |S_2\rangle + \ldots + |S_N\rangle)^{\eta}$.*

(iv)    Make a measurement that projects each subsystem onto one of its basis states that points to an item in the database. The item that the most subsystems point to, is the marked item.

*Since the probability of obtaining the basis state corresponding to the marked item ($S_1$) in each of the $\eta$ subsystems is approximately $\frac{9}{N}$ and the probability of obtaining another basis state is approximately $\frac{1}{N}$, it follows by the law of large numbers* [Feller], *that out of $\eta$ subsystems, $\frac{9\eta}{N} \pm O\left(\sqrt{\frac{\eta}{N}}\right)$ lie in state $S_1$ while $\frac{\eta}{N} \pm O\left(\sqrt{\frac{\eta}{N}}\right)$ lie in each of the other basis states. If $\eta = KN$, then $9K \pm O(\sqrt{K})$ subsystems lie in $S_1$ and $K \pm O(\sqrt{K})$ in each of the other basis states. If $K \gg 1$, then the uncertainty due to the $\pm O(\sqrt{K})$ term can be neglected when compared to the dominant term that is proportional to $K$.*

*In fact, it follows by the central limit theorem* [Feller], *that the probability of a particular variable deviating by more than $\pm \gamma \sqrt{K}$ from its expected value is less than $\exp(-\Omega(\gamma^2))$. Therefore if $K = \Omega(\log N)$ (equivalently if $\eta = \Omega(N \log N)$), then with a probability approaching unity, $S_1$ occurs with a frequency greater than any of the $(N-1)$ other basis states.*

## 4. Discussion

(i) The architecture of a system that does the above calculation would be something like that in the schematic below:

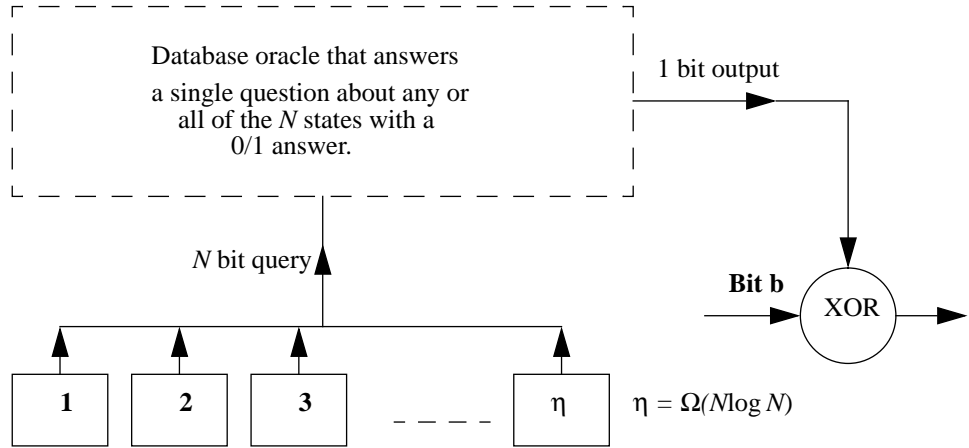

**$\eta$ identical subsystems - each subsystem is placed in a superposition of $N$ states with equal amplitudes. Bit b is placed in the superposition ($|0\rangle - |1\rangle$) ; as mentioned in section 2, this configuration inverts the phase of all states for which the 1 bit output from the oracle is 1.**

The $N$ bit query to the oracle in the above figure is: "Is the total number of marked items corresponding to all the subsystems even or odd?" Note that an $N$ bit query is required for this. For example, let there be 4 items, i.e. $N = 4$, denoted by *A, B, C & D*. $\eta$ is larger than $N$ by a factor of $\Omega(\log N)$, assume $\eta = 20$. Assume that *C* is the marked item. The system (composed of $\eta$ subsystems) is in a superposition of states, each element of this superposition corresponds to a particular set of items as in step (i) of section 3. For example, in one element of the superposition, we may have (out of $\eta = 20$ subsystems), *2* in a basis state corresponding to item *A*, *5* corresponding to item *B*, *8* corresponding to item *C* & *5* to item *D*. The query to the oracle would be the 4 bit query: 0101, denoting that the first & third items (*A & C*) have an even number of subsystems in their corresponding basis states (hence the first & third bits of the query are 0's) and the second and fourth (*B & D*) have an odd number of subsystems (the second & fourth bits are *1*'s). The oracle, knowing that the marked item was *C,* would look at the third bit of the query (*0*) and return the one bit answer (1), i.e. the marked item's basis state is present an even number of times. In the terminology of section 2, the function $f(\bar{x})$ that the oracle evaluates, is 1 if the query indicates that the marked item's basis state is present an odd number of times, 0 otherwise. By having bit *b* (shown in figure above) in the superposition $\frac{1}{\sqrt{2}}(|0\rangle - |1\rangle)$, this operation selectively inverts the phase of the amplitude in the basis state corresponding to the marked item for *each* subsystem.

The *inversion about average* operation increases the amplitude in the basis state corresponding to the marked item ($C$) in all of the 20 subsystems. Finally a measurement is made which projects each subsystem into one of its basis states, the basis state corresponding to the marked item ($C$) has a higher probability of occurring. Since the same difference in probability occurs in each of the subsystems, it follows that by choosing the number of subsystems to be sufficiently large ($\eta = 20$), this small difference in probabilities can be detected.

(ii) As mentioned previously, the same algorithm applies when more than one item is marked, with the caveat that the number of marked items is less than $\frac{N}{4}$. There are two reasons for this limitation.

First, there is no way of distinguishing the cases when $k$ items were marked or when $(N - k)$ items were marked. This is because if $k$ items are marked, then after step (ii), the state vector is of the form $(-|S_1\rangle - |S_2\rangle - \ldots - |S_k\rangle + |R_1\rangle + |R_2\rangle + \ldots |R_{N-k}\rangle)^\eta$ where the $k$ items corresponding to the $S$ states are marked and those corresponding to the $R$ states are not. This is indistinguishable from the state vector $(|S_1\rangle + |S_2\rangle + \ldots + |S_k\rangle - |R_1\rangle - |R_2\rangle - \ldots |R_{N-k}\rangle)^\eta$ which is obtained if the $(N - k)$ items correponding to the $R$ states were marked.

The second reason is that, when the number of marked items approaches $\frac{N}{2}$, the difference of probabilities that needs to be resolved is very small and it needs more than $\Omega(N \log N)$ subsystems to do this. In the terminology of the previous paragraph, this happens because $k$ becomes very close to $(N - k)$.

(iii) An argument, sometimes quoted, is that since a quantum mechanical system needs at least $\Omega(\sqrt{N})$ steps in order to identify a marked item out of $N$ possible items [BBBV] & [BBHT], it could not possibly solve an NP-complete problem in polynomial time (since an NP-complete problem has an exponential number of items). This paper demonstrates that it is possible to overcome this particular $\Omega(\sqrt{N})$ bottleneck by having more elaborate queries. However, even though there is just a single query, the preprocessing and postprocessing steps required are still $\Omega(N \log N)$.

## 5. Acknowledgments

Would like to thank Asher Peres, Peter Hoyer, Dan Gordon, Anirvan Sengupta and most of all, Norm Margolus for their timely feedback.